\documentstyle[twocolumn,aps,prl]{revtex}
\begin{document}
%\twocolumn[\hsize\textwidth\columnwidth\hsize\csname @twocolumnfalse\endcsname]
\title{Energy spectra of cosmic rays accelerated at   
ultrarelativistic shock waves}

\author{ J. Bednarz and M. Ostrowski}
\address{ Obserwatorium Astronomiczne, Uniwersytet Jagiello\'nski, 
ul Orla 171, 30-244 Krak\'ow, Poland }

\date{\today}
\maketitle 
\begin{abstract}
Energy spectra of particles accelerated by the first-order Fermi   
mechanism are investigated at ultrarelativistic shock waves, outside 
the range of Lorentz factors considered previously. For particle
transport near the shock a numerical method involving small amplitude
pitch-angle scattering is applied for flows with Lorentz factors
$\gamma$ from 3 to 243. For large $\gamma$ shocks a convergence of
derived energy spectral indices up to the value $\sigma_\infty \approx 2.2$ 
is observed for all considered turbulence amplitudes and magnetic field 
configurations. Recently the same index was derived for $\gamma$-ray
bursts by Waxman [Astrophys. J. Lett. {\bf 485}, L5 (1997)].
\end{abstract}

%\pacs{98.70.Sa, 98.70.Rz}
\vskip1.2pc

In currently favored gamma-ray burst (GRB) models optically thin 
emitting regions move relativistically, with Lorentz factors of order
of a few hundreds (cf. a review [1]). The power-law form of the spectrum
often observed at high photon energies suggests the existence of
nonthermal population of energetic particles. It was also proposed
that GRB sources may produce cosmic ray particles with extremely high
energies [2]. Thus modeling of burst sources requires a discussion
of particle acceleration processes, possibly the Fermi acceleration
at ultrarelativistic shock waves.
 
The work of Kirk and Schneider [3] opened the problem of cosmic ray
acceleration at relativistic shock waves for quantitative 
consideration. Substantial progress since that time (e.g. Heavens and
Drury [4], Kirk and Heavens [5], Begelman and Kirk [6], Ostrowski [7],
Bednarz and Ostrowski [8]; for a review see Ostrowski [9] and Kirk [10])
clarified a number of issues related to shock waves with velocities
reaching $0.98 c$ or the Lorentz factor $\gamma \simeq 5$ , but -- to
our knowledge -- no one has attempted to discuss particle acceleration at
shocks moving with ultrarelativistic velocities characterized with
large factors $\gamma >> 1$.
    
The main difficulty in modelling an acceleration process at shocks with 
large $\gamma$ is the fact that involved particle distributions are
extremely anisotropic in shock, with the particle angular
distribution opening angles $\sim \gamma^{-1}$ in the upstream plasma
rest frame. When transmitted downstream the shock particles have a
limited chance to be scattered so efficiently to reach the shock again,
but the energy gain of any such ``successful'' particle can be comparable
to its original energy. As pointed out by Bednarz and Ostrowski [8] any
realistic model of particle scattering at magnetohydrodynamic turbulence
close to the relativistic shock cannot involve large-angle pointlike
scattering. The choice is either to integrate exactly particle equations
of motion in some ``realistic'' structure of the perturbed magnetic field,
or to use a small-angle scattering model for particle momentum. With the
angular scattering amplitude $\Delta \Omega << \gamma^{-1}$ and the mean
scattering time $\Delta t$ not too short [$\Delta t \ge T_g \, (\Delta
\Omega)^2$ , where $T_g$ is the particle gyration period], the last
model reproduces the pitch-angle diffusion process at small amplitude
waves. We prefer that approach to the exact integration of equations
of motion of a particle because of its relative simplicity. It is
also suggested that it can be reasonably used for modeling particle
trajectories in turbulent fields with large amplitude, if small $\Delta t$
[$\le T_g \, (\Delta \Omega)^2$] is involved. Below, a hybrid method
involving the small amplitude pitch-angle scattering is applied for
a particle transport near the shock for flows with Lorentz factors
$\gamma$ from 3 to 243.

{\it I. Numerical simulations. -- }
In the present considerations we model a process of cosmic ray particle
acceleration by applying the following Monte Carlo approach (cf. e.g.
[8]). Energetic seed particles are injected at the shock and each
particle trajectory is followed using numerical computations until it
escapes through the free escape boundary placed far downstream from
the shock or it reaches the energy larger than the upper energy limit at
$E_{max} = 10^{10} E_0$ ($E_0$ - initial energy). Simulations 
are continued until one obtains the power-law spectrum in the full
range ($E_0$, $E_{max}$) (cf. [7,8]). All computations are performed
in the respective - upstream or downstream - plasma rest frame. Each
time when the particle crosses the shock its momentum is Lorentz transformed
to the respective plasma rest frame and, in the shock normal rest frame
(cf. [6]; henceforth ``shock rest frame'') the respective contribution is
added to the given momentum bin in the particle spectrum. In the
simulations we use a simple trajectory splitting technique. All
particles are injected into simulations with the same initial weight
factors $1.0$~. When some particles escape through the boundary,
we replace them with ones arising from splitting the remaining high-weight
particles in a way to preserve its phase space coordinates but
ascribing only a half or a smaller respective part of the original
particle weight to each of the resulting particles. For any particle
crossing the shock a factor is added to the simulated spectrum in the
shock rest frame equal to the particle weight divided by its velocity
component normal to the shock.
 
Efficient particle scattering with a very small $\Delta \Omega$ requires
derivation of a large number of scattering acts and the respective
numerical code becomes extremely time-consuming. In order to overcome
this difficulty in the present simulations we propose a hybrid approach
involving very small $\Delta \Omega_1$ ($<< \gamma^{-1}$) close to the
shock, where the scattering details play a role, and much larger
scattering amplitude $\Delta \Omega_2 = 9^\circ$ to describe particle
diffusion further away from the shock. The respective
scaling of the scattering time $\Delta t$ is performed in both cases
($\Delta \Omega_1^2/\Delta t_1=\Delta \Omega_2^2/\Delta t_2$) to mimic
the same turbulence amplitudes measured by the values of the cross-field
diffusion coefficient, $\kappa_\perp$, and the parallel diffusion
coefficient, $\kappa_\|$.

For a few instances we checked the validity of this approach by
reproducing the results for the small $\Delta \Omega_1$ everywhere. In
the present simulations we assume the same scattering conditions upstream
and downstream from the shock (the same $\kappa_\perp$ and $\kappa_\|$ in
the units of $r_g c$, where $r_g c$ is the particle gyration radius in the
unperturbed background magnetic field), preserving particle energy at
each scattering in the plasma rest frame. In the simulations we
considered a few configurations of the upstream magnetic field, with
inclinations with respect to the shock normal being $\psi = 0^\circ$,
$10^\circ$,$20^\circ$, $30^\circ$,$60^\circ$ and $90^\circ$. The first
case represents the parallel shock, the second is for the oblique shock
- subluminal (i.e. with the shock velocity projection at the magnetic
field with slower than light velocity) at $\gamma = 3$ and a superluminal
one at larger $\gamma$, and the larger $\psi$ are for
superluminal perpendicular shocks for all velocities. The downstream
magnetic field is derived for the relativistic shock with the
compression $R$ obtained with the formulas of Heavens and Drury [4] for
a cold ($e$, $p$) plasma -- $R \approx 3.6$ for our smallest value of
$\gamma = 3$ and tends to $R = 3$ for $\gamma >> 1$, as measured in the
shock rest frame.

{\it II. Results. -- }
Particle spectral indices were derived for different mean
magnetic field configurations, measured by the magnetic field
inclination $\psi$ with respect to the shock normal in the upstream
plasma rest frame, and for different amounts of turbulence measured by
$\kappa_\perp / \kappa_\|$.

In successive panels in Fig.~1 the energy spectral indices, $\sigma$,
for varying $\psi$ and $\kappa_\perp/ \kappa_\|$ are presented. For a
parallel shock ($ \psi = 0^\circ $) the amount of scattering does not
influence the spectral index and for growing $\gamma$ it approaches
$\sigma_\infty \simeq 2.2$. One may note that essentially the same
limiting value was anticipated for the large-$\gamma$ parallel shocks
by Heavens and Drury [4]. The results for $\psi = 10^\circ$ are for
superluminal shocks if $\gamma > 5.75$. 

In this case, when we go from the ``slow'' $\gamma = 3$ shocks to higher
$\gamma$ ones, at first the spectrum inclination increases ($\sigma$
grows), but at large $\gamma$ the spectrum flattens to approach the
asymptotic value close to 2.2. The spectrum steepening phase is more
pronounced for small amplitude perturbations 
(small $\kappa_\perp / \kappa_\|$), but even at very low turbulence
levels the final range of the spectrum flattening is observed.
For
larger $\psi$ the situation does not change considerably, but the phase
of spectrum steepening is wider, involving larger values of $\sigma$
and starting at smaller velocities, below the lower limit of our
considerations (there may be no such range involving the small
\begin{figure}
\vspace*{11.5cm}
%\special{psfile=f1.ps hscale=49.0 vscale=49.0 hoffset=104 voffset=-40}
\includegraphics{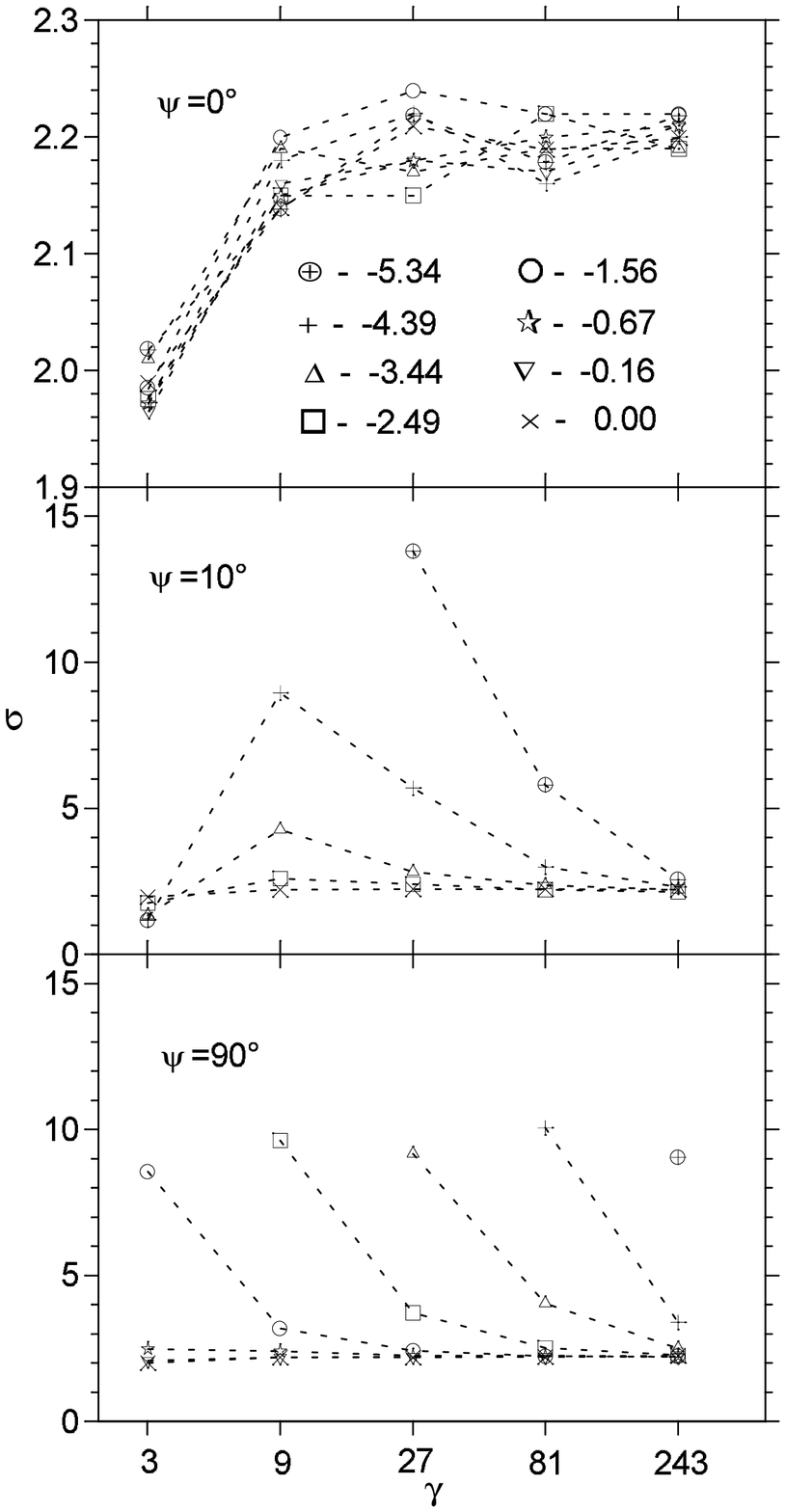}
\caption{   
The simulated spectral indices $\sigma$ for particles accelerated at   
shocks with different Lorentz factors $\gamma$. Results for a given   
$\kappa_\perp / \kappa_\| $ are joined with lines; the respective value   
of $\log_{10} \kappa_\perp / \kappa_\| $ is marked by the point shape (see
upper panel). The results for different magnetic field inclinations
$\psi$ are given in the successive panels: (a) $\psi = 0^\circ$,
(b) $\psi = 10^\circ$, and (c) $\psi = 90^\circ$.}
\label{fig1}
\end{figure}
\noindent
steepening phase if the required velocity is below the sound velocity).
The spectral indices for different magnetic field inclinations, but for
the same value of $\log_{10} ( \kappa_\perp / \kappa_\| ) = -3.44$, are
presented at Fig.~2.

The large spectral indices occurring in the steepening phase are
usually interpreted as a spectrum cutoff. In this case the main
factor increasing the particle energy density is a  nonadiabatic
compression in the shock [6].
The particle angular distributions $F(\mu)$ in the considered shocks can
be extremely anisotropic when considered in the upstream plasma rest
frame. However, when presented in the shock rest frame the distribution
is always ``mildly'' anisotropic. This feature is illustrated at Fig.~3
for $\gamma$ equals 3 or 27 (note that in Figs.~3-6 the area below
each curve is normalized to 100). In simulations we observed an
interesting phenomenon accompanying previously discussed spectrum
convergence to the limiting inclination: spectra close to the limit
exhibit similar angular distributions at the shock {\it as measured in
the shock}
\begin{figure}
\vspace*{5.5cm}
%\special{psfile=f2.ps hscale=50.0 vscale=50.0 hoffset=42 voffset=235}
\includegraphics{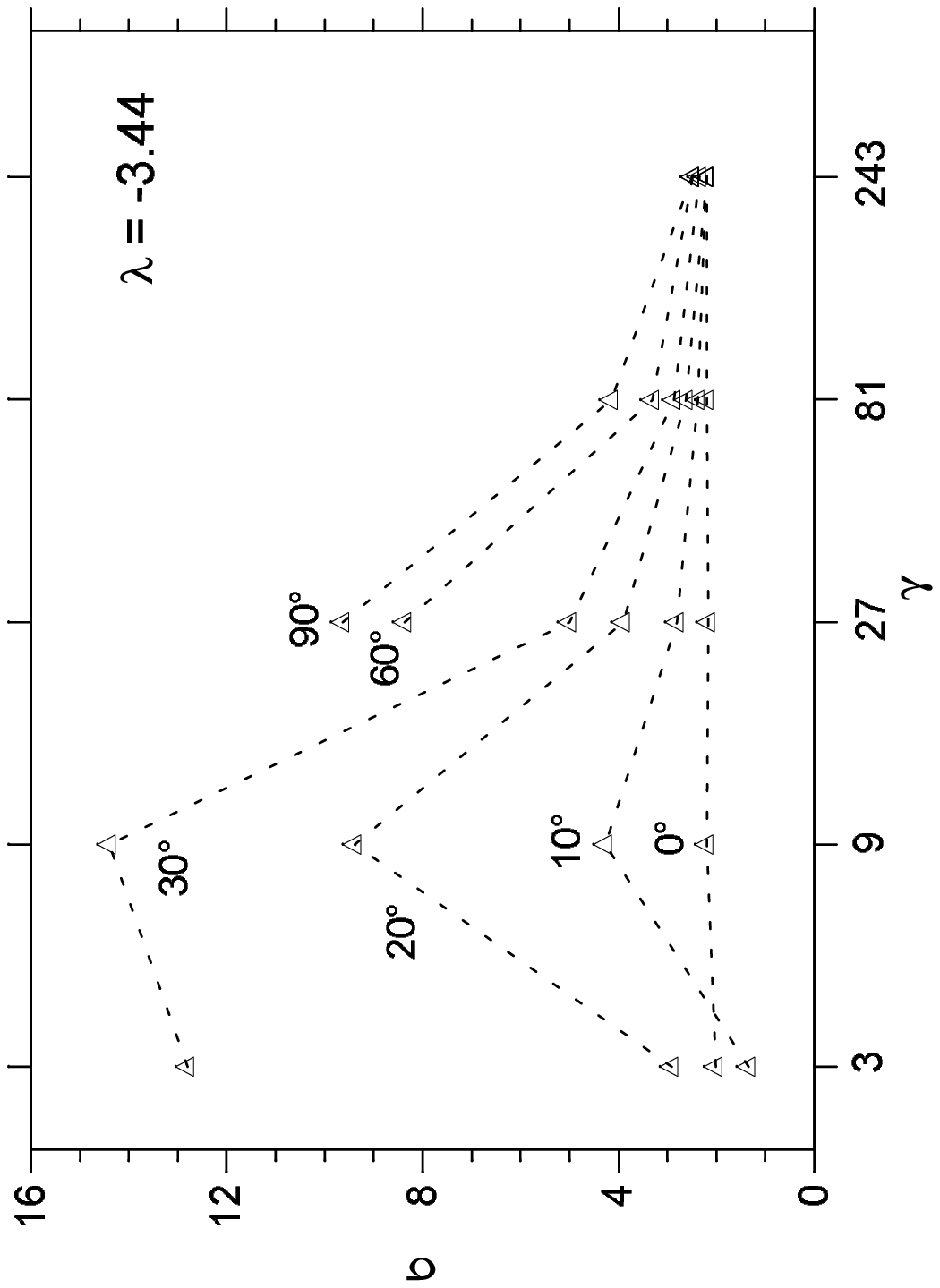}
\caption{
The simulated spectral indices $\sigma$ for particles accelerated at   
shocks with different Lorentz factors $\gamma$. Results for a given   
upstream magnetic field inclination $\psi$ are joined with dashed   
lines; the respective value of $\psi$ is given near each curve. The   
value $\lambda \equiv \log_{10} (\kappa_\perp / \kappa_\| )$ is given in the   
figure.}
\label{fig2}
\end{figure}
\begin{figure}
\vspace*{11.5cm}
%\special{psfile=f3.ps hscale=49.0 vscale=49.0 hoffset=104 voffset=-40}
\includegraphics{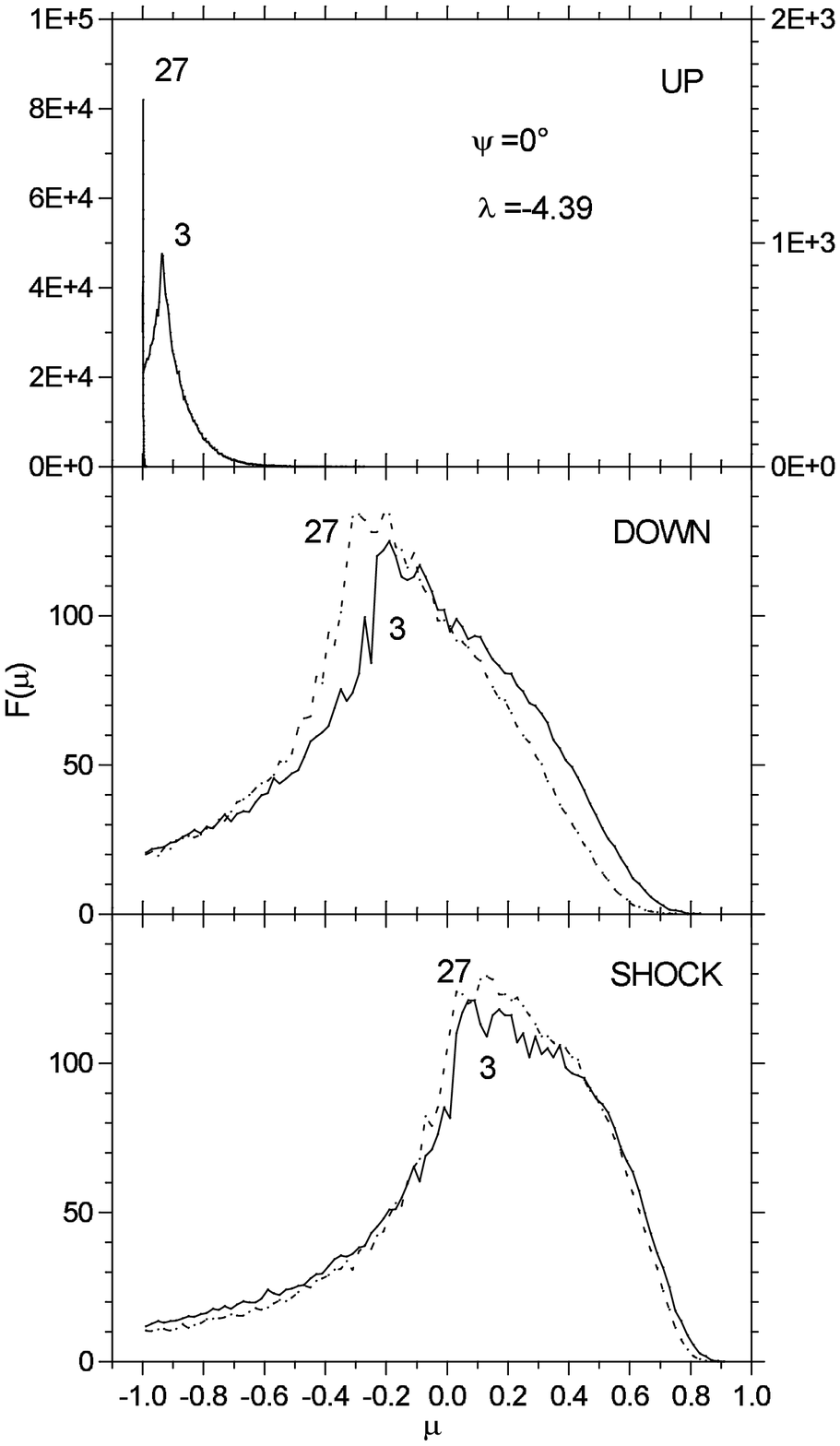}
\caption{
The simulated particle angular distributions in the shock in different
coordinate frames: UP: the upstream plasma rest frame, DOWN: the
downstream plasma rest frame, and SHOCK: the shock rest frame. The
results are presented for parallel shocks with the Lorentz factors
27 and 3 given near the respective curves. In the upper panel the left
axis is for $\gamma = 27$ and the right one for $\gamma = 3$.}
\label{fig3}
\end{figure}
\noindent
{\it rest frame} (Fig.~4). Again, this feature is independent of
the background conditions, and the difference between the actual angular
distribution and the limiting one reflects the difference between the
spectral index $\sigma$ and $\sigma_\infty$~(cf. Fig.~5). For parallel
shocks with $\gamma \ge 9$, where the spectral index is essentially
constant $\sigma = \sigma_\infty$, this distribution is independent of
the value of $\gamma$ and the perturbation amplitude $\kappa_\perp /
\kappa_\|$ (Fig.~6).

{\it III. Discussion. -- }
For large $\gamma$ shocks we observe the convergence of the derived
energy spectral indices to the value $\sigma_\infty \approx 2.2$,
independently of background conditions. This unexpected result providing
a strong constraint for the acceleration process in large $\gamma$ shocks
requires a more detailed analysis. Particularly, the acceleration
process in the superluminal shocks has to be clarified in the presence
of the small amplitude turbulence.
    
The inspection of particle trajectories reveals a simple picture of
acceleration. Cosmic ray particles are wandering in the downstream
region with the shock wave moving away with the mildly relativistic
velocity $ \approx c/3$. Some of these particles succeed to reach the
shock, but then they remain in the upstream region for a very short time
-- being very close to the shock -- due to large shock velocity $\approx
c$. This scenario is essentially equivalent to the picture involving
particles reflecting in a nonelastic way from the receding wall.

For large $\gamma$ shocks any particle crossing the shock upstream has
a momentum vector nearly parallel to the shock normal (cf. Ostrowski
[7]); e.g., for $\gamma = 243$ the momentum inclination must be smaller
than $\theta_{max} \approx 0.24^\circ$.
If the scattering or the movement along the curved trajectory increases
this inclination above the mentioned limiting value the particle tends
to recross the shock downstream.
One should note that even a tiny -- comparable to
$\theta_{max}$ -- angular deviation in the upstream plasma ($\Delta
\theta_{U}$) can lead to large angular deviation for $\gamma \gg 1$ as
observed in the downstream rest frame.
The phenomenon of decreasing
$\sigma$ to $\sigma_\infty$ at constant $\lambda$ and for growing
$\gamma$ results from {\em slower diminishing
of the part of $\Delta
\theta_{U}$ caused by scattering in comparison to $\Delta \theta_U$
arising due to trajectory curvature} in the uniform field compo-
\begin{figure}
\vspace*{5.5cm}
%\special{psfile=f4.ps hscale=50.0 vscale=50.0 hoffset=42 voffset=235}
\includegraphics{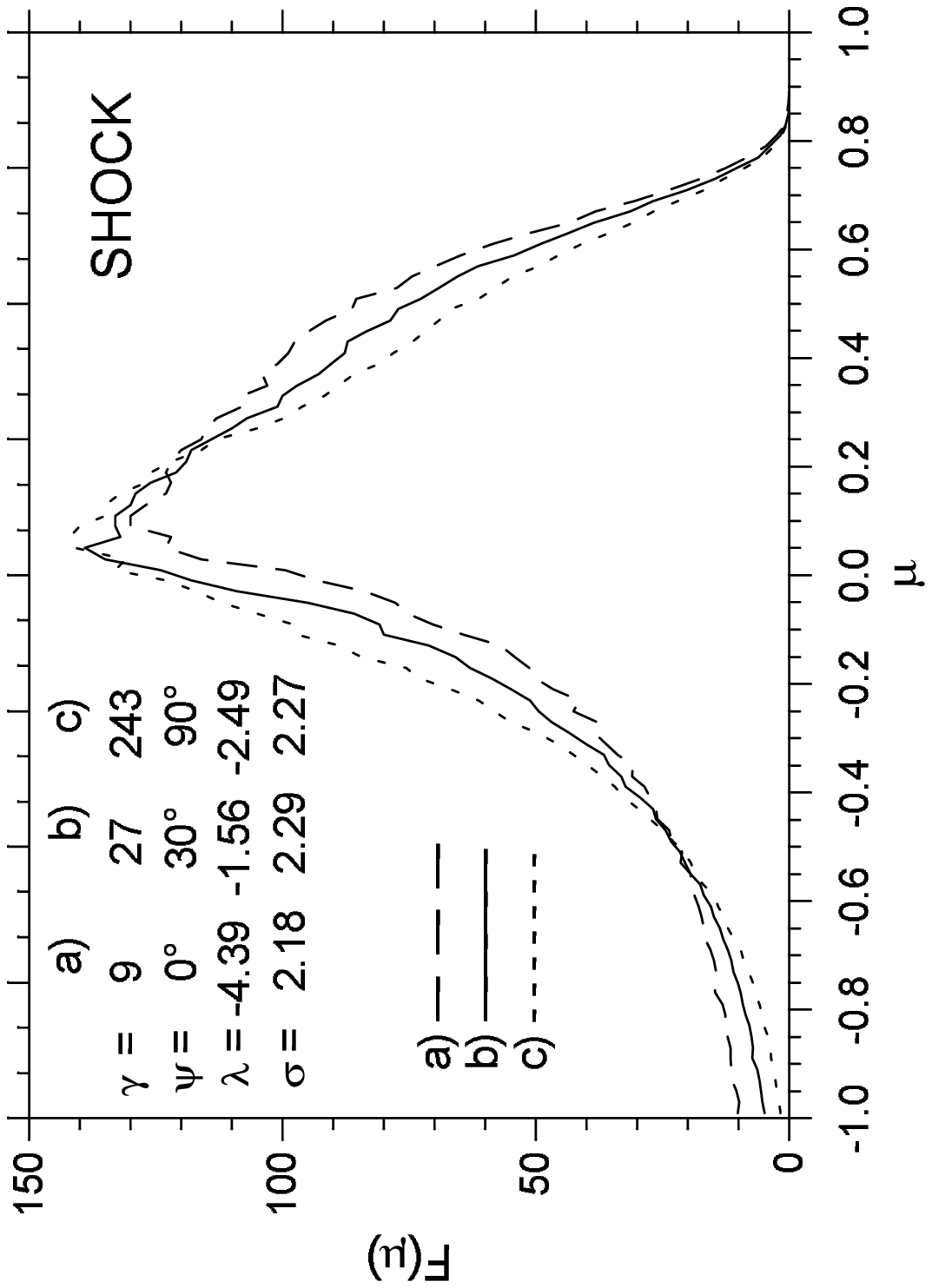}
\caption{
Examples of the shock rest frame  particle angular distributions 
for different cases with $\sigma$ close to $\sigma_\infty$.}
\label{fig4}
\end{figure}
\begin{figure}
\vspace*{5.5cm}
%\special{psfile=f5.ps hscale=50.0 vscale=50.0 hoffset=42 voffset=235}
\includegraphics{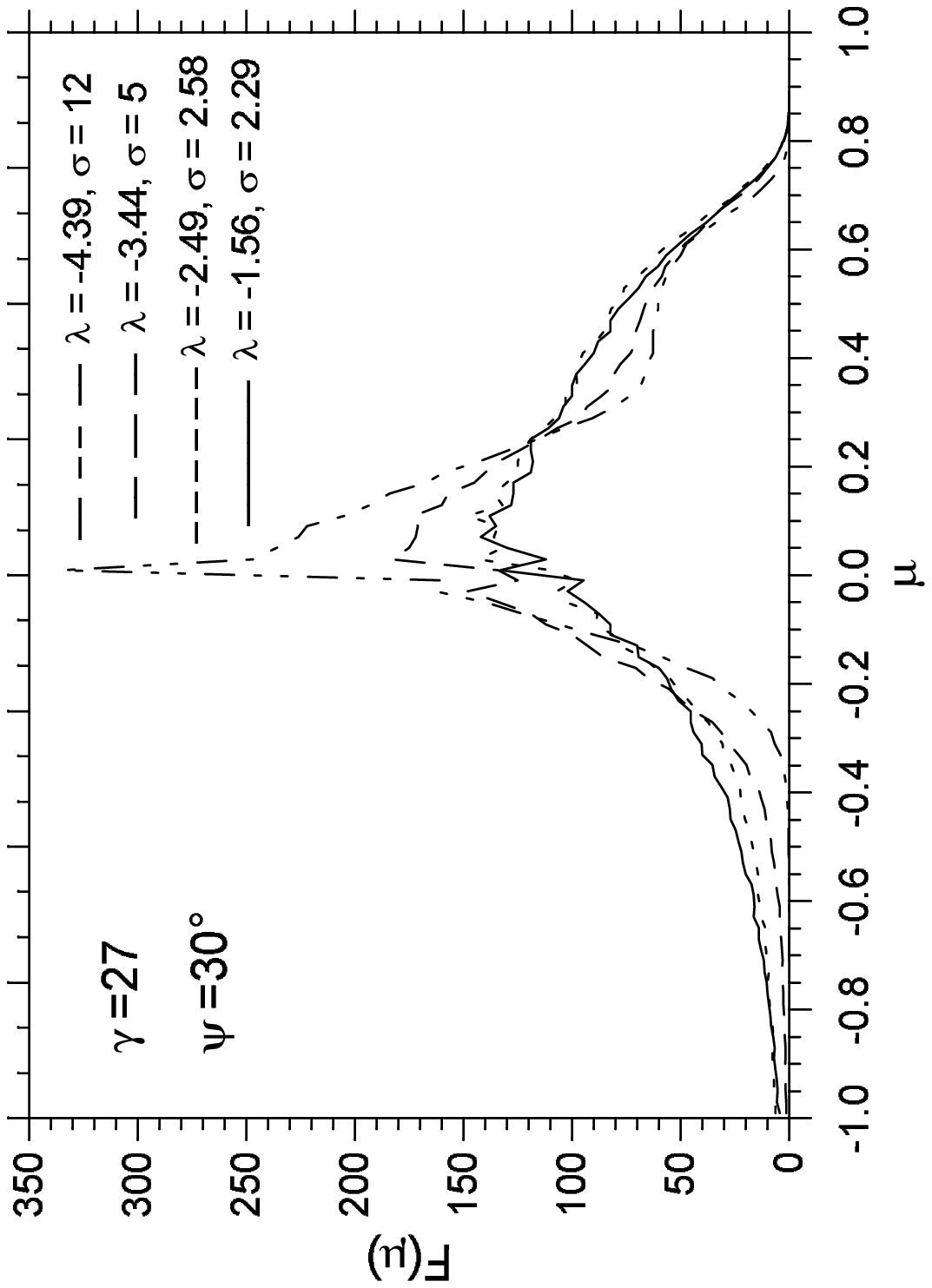}
\caption{
The shock rest frame  particle angular distributions for $\gamma = 
27$ and  $\psi = 30^\circ$. Curves are presented for increasing 
$\lambda \equiv \log_{10} \kappa_\perp / \kappa_\| $ and $\sigma$
approaching $\sigma_\infty$. The last curve is the same as the curve
(b) at Fig. 4.}
\label{fig5}
\end{figure}
\noindent
nent. In this way the magnetic field structure defined by $\psi$
becomes unimportant, at least for $\gamma \to \infty$ (one should 
also note that the downstream field inclination approaches $90^\circ$
if $\psi\neq 0$ and $\gamma \gg 1$). As a result particles crossing
the shock downstream are scattered in a wide angular range with
respect to the shock normal, providing some particles with the
trajectory phase parameter allowing for recrossing the shock upstream
even for the perpendicular magnetic field configuration. Our
interesting finding not fully explained with such simple arguments is
of the belief that the resulting spectral index is the same for
oblique and  parallel shocks.
We also observe that when approaching the limiting value of the
spectral index the mean particle energy gain $<\Delta E/E>$ in the
cycle ``upstream-downstream-upstream'' reaches a value close
(slightly above) to $1.0$, much smaller than a factor $\gamma^2$
expected for a model involving a large angle
\begin{figure}
\vspace*{5.5cm}
%\special{psfile=f6.ps hscale=50.0 vscale=50.0 hoffset=42 voffset=235}
\includegraphics{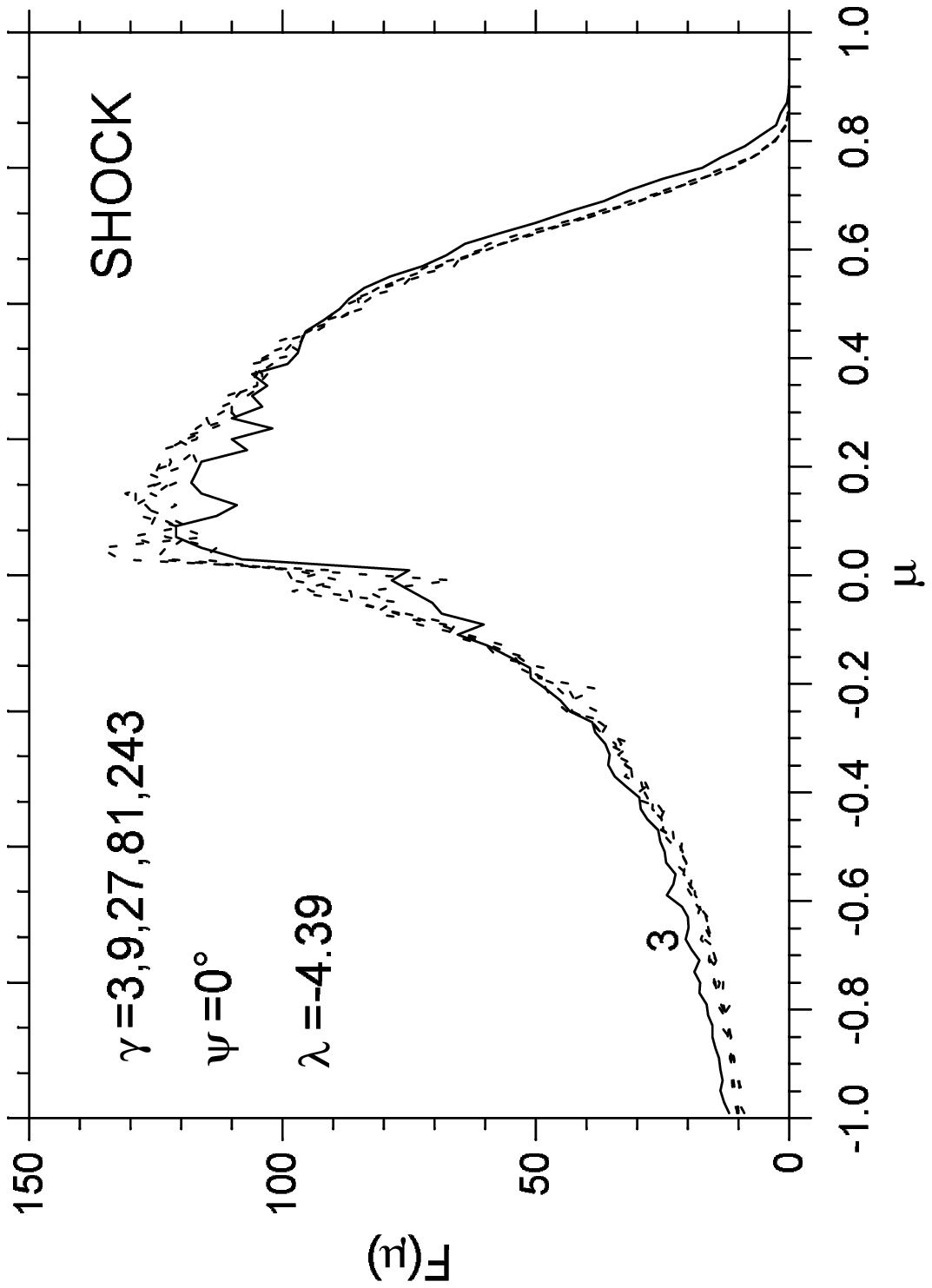}
\caption{
The shock rest frame  particle angular distributions for
parallel shocks with $\gamma = 3$, $9$, $27$, $81$, $243$.
A visible deviation of the distribution for $\gamma = 3$ (full line)
results from the slightly larger compression occurring in such a shock.}
\label{fig6}
\end{figure}
\noindent
pointlike scattering. Thus the particle acceleration time scale, as measured
in the downstream plasma rest frame, can be roughly estimated as a
fraction of the gyration time in this region.

{\it IV. Final remarks. -- }
The presented results are to be applied in models of GRB sources
involving ultrarelativistic shock waves. One should note that the
mean downstream plasma proton energies can reach there several tens
of GeV (cf. Paczy\'nski and Xu [11]) and the lower limit of the
considered cosmic ray energies has to be larger than this scale.
For shocks propagating in (e$^-$, e$^+$) plasma the involved thermal
energies are lower, $\sim \gamma$ MeV. These estimates provide
the respective lower limits for the accelerated cosmic ray particles.
For the physical conditions considered in GRB sources the
acceleration process can provide particles with much larger energies,
limited only by the condition that the energy loss processes
(radiative, or due to escape) are ineffective in the downstream
gyroperiod time scale. We note a striking coincidence of our limiting
spectral index with the value $2.3$ $\pm 0.1$ derived for energetic
electrons from gamma-burst afterglow observations [12]. 
 
Our derivations are limited to the test particle approach. However,
as the obtained spectra are characterized with $\sigma > 2.0$, any 
nonlinear back reaction effects are not expected to affect the
acceleration process within the spectrum high energy tail with
$\sigma \approx \sigma_\infty$.
   
The present work was supported by the {\it Komitet Bada\'n Naukowych}
through the grant PB 179/P03/96/11.

\end{document}